\documentclass[reprint,showpacs,aps,prd,nofootinbib,superscriptaddress,longbibliography]{revtex4-1}

\usepackage{graphicx}
\usepackage[utf8]{inputenc} 
\usepackage{placeins}
\usepackage{amsmath}
\usepackage{amssymb}
\usepackage{float}
\usepackage{xcolor,color,colortbl}
\usepackage{comment}
\usepackage{multirow}
\usepackage{color}
\usepackage{soul}

\def\beq{\begin{equation}}
\def\eeq{\end{equation}}
\def\bea{\begin{eqnarray}}
\def\eea{\end{eqnarray}}

\usepackage{amsbsy}
\usepackage{bm}
\usepackage{color}
\usepackage{fancyhdr}
\usepackage[pdftex]{epsfig}
\usepackage[colorlinks=true,linkcolor=blue,citecolor=blue,urlcolor=magenta,filecolor=blue]{hyperref}
\usepackage{slashed}

\begin{document}

\bigskip

\vspace{2cm}

\title{Extra gauge bosons and lepton flavor universality violation in $\Upsilon$ and $B$ meson decays}

\vskip 6ex

\author{Cristian H. Garc\'{i}a-Duque}
\email{chgarcia@uniquindio.edu.co}
\affiliation{Programa de F\'{i}sica, Universidad del Quind\'{i}o, Carrera 15 Calle 12 Norte, C\'{o}digo Postal 630004, Armenia, Colombia}
\author{J. H. Mu\~{n}oz}
\email{jhmunoz@ut.edu.co}
\affiliation{Departamento de F\'{i}sica, Universidad del Tolima, C\'{o}digo Postal 730006299, Ibagu\'{e}, Colombia}
\author{N\'{e}stor Quintero}
\email[]{nestor.quintero01@usc.edu.co}
\thanks{(Corresponding author)}
\affiliation{Facultad de Ciencias B\'{a}sicas, Universidad Santiago de Cali, Campus Pampalinda, Calle 5 No. 62-00, C\'{o}digo Postal 76001, Santiago de Cali, Colombia}
\affiliation{Departamento de F\'{i}sica, Universidad del Tolima, C\'{o}digo Postal 730006299, Ibagu\'{e}, Colombia}
\author{Eduardo Rojas}
\email{eduro4000@gmail.com}
\affiliation{Departamento de Física, Universidad de Nari\~no, A.A. 1175, San Juan de Pasto, Colombia}

\bigskip

\begin{abstract}
Lepton flavor universality can be tested through the ratio of semileptonic $B$ meson decays and leptonic $\Upsilon$ meson decays, with $\Upsilon \equiv \Upsilon(nS)$ ($n=1,2,3$). For the charged-current transitions $b \to c\tau\bar{\nu}_\tau$, discrepancies between the experiment and the Standard Model (SM) have been observed in recent years by different flavor facilities such as BABAR, Belle, and LHCb. While for the neutral-current transitions $b \bar{b} \to \tau\bar{\tau}$, the BABAR experiment reported recently a new measurement of leptonic decay ratio $R_{\Upsilon(3S)} = {\rm BR}(\Upsilon(3S) \to \tau^+\tau^-)/{\rm BR}(\Upsilon(3S) \to \mu^+\mu^-)$, showing an agreement with the SM at the $1.8 \sigma$ level. In light of this new BABAR result and regarding the connection between new physics (NP) interpretations to the charged-current $b \to c \tau \bar{\nu}_{\tau}$ anomalies and neutral-current $b  \bar{b} \to \tau \bar{\tau}$ processes, in this study, we revisit the NP consequences of this measurement within a simplified model with extra massive gauge bosons that coupled predominantly to left-handed leptons of the third-generation. We show that the BABAR measurement of $R_{\Upsilon(3S)}$ cannot easily be accommodated (within its experimental $1\sigma$ range) together with the other $b \to c\tau\bar{\nu}_\tau$ data, hinting toward a new anomalous observable. 
\end{abstract}

\maketitle

\section{Introduction}

In recent years, tantalizing hints of lepton flavor universality (LFU) violation have been suggested by the experiments BABAR, Belle, and LHCb in the measurements of the ratio of semileptonic $B$ meson decays~\cite{Lees:2012xj,Lees:2013uzd,Huschle:2015rga,Sato:2016svk,Hirose:2017vbz,Aaij:2015yra,Aaij:2017deq,
Aaij:2017uff,Belle:2019rba,Hirose:2017dxl,Hirose:2016wfn,Amhis:2019ckw,
HFLAVsummer,Aaij:2017tyk}
\begin{eqnarray}
R(D^{(\ast)}) &=& \dfrac{{\rm BR}(B \to D^{(\ast)}\tau \bar{\nu}_\tau)}{{\rm BR}(B \to D^{(\ast)}\ell^\prime \bar{\nu}_{\ell^\prime})}  \ \ \ (\ell^\prime = e \ {\rm or} \ \mu), \label{R_D} \\
R(J/\psi) &=&  \dfrac{{\rm BR}(B_c \to J/\psi \tau \bar{\nu}_\tau)}{{\rm BR}(B_c \to J/\psi\mu \bar{\nu}_{\mu})}. \label{R_Jpsi} 
\end{eqnarray} 

\noindent The latest 2019 world averages values reported by the Heavy Flavor Averaging Group (HFLAV) on the measurements of $R(D^{(\ast)})$~\cite{Amhis:2019ckw,HFLAVsummer} and the LHCb results on $R(J/\psi)$~\cite{Aaij:2017tyk,Harrison:2020nrv,Harrison:2020gvo},
\begin{eqnarray}
R(D) &=&
\begin{cases}
\text{HFLAV:} \ 0.340 \pm 0.027 \pm 0.013\text{~\cite{Amhis:2019ckw,HFLAVsummer}}, \\
\text{SM:} \ 0.299 \pm 0.003\text{~\cite{Amhis:2019ckw,HFLAVsummer}},
\end{cases} \\
R(D^\ast) &=&
\begin{cases}
\text{HFLAV:} \ 0.295 \pm 0.011 \pm 0.008\text{~\cite{Amhis:2019ckw,HFLAVsummer}}, \\
\text{SM:} \ 0.258 \pm 0.005\text{~\cite{Amhis:2019ckw,HFLAVsummer}},
\end{cases} \\
R(J/\psi) &=& \begin{cases}
\text{LHCb:} \ 0.71 \pm 0.17 \pm 0.18\text{~\cite{Aaij:2017tyk}}, \\
\text{SM:} \ 0.283 \pm 0.048\text{~\cite{Harrison:2020nrv,Harrison:2020gvo}},
\end{cases} 
\end{eqnarray}

\noindent exhibit a deviation with respect to the Standard Model (SM) expectations by $1.4\sigma$, $2.5\sigma$, and $1.8\sigma$, respectively. SM predictions for $R(D^{(\ast)})$ are taken from the average values obtained by HFLAV~\cite{HFLAVsummer}, while for $R(J/\psi)$ we consider the recent lattice QCD calculations~\cite{Harrison:2020nrv,Harrison:2020gvo}. Furthermore, polarization observables such as the $\tau$ lepton polarization $P_\tau(D^\ast)$ and the longitudinal polarization of the $D^*$ meson $F_L(D^\ast)$ related with the channel $\bar{B} \to D^\ast \tau \bar{\nu}_\tau$ have been observed in the Belle experiment~\cite{Hirose:2017dxl,Hirose:2016wfn,Abdesselam:2019wbt},
\begin{eqnarray}
P_\tau(D^\ast) &=&
\begin{cases}
\text{Belle:} \ -0.38 \pm 0.51 {}^{+0.21}_{-0.16}\text{~\cite{Hirose:2017dxl,Hirose:2016wfn}}, \\
\text{SM:} \ -0.497 \pm 0.013\text{~\cite{Tanaka:2012nw}},
\end{cases} \\
F_L(D^\ast) &=&
\begin{cases}
\text{Belle:} \ 0.60 \pm 0.08 \pm 0.035\text{~\cite{Abdesselam:2019wbt}}, \\
\text{SM:} \ 0.46 \pm 0.04\text{~\cite{Alok:2016qyh}},
\end{cases} 
\end{eqnarray}

\noindent and also present a disagreement respect with the corresponding SM predictions~\cite{Tanaka:2012nw,Alok:2016qyh}. Additionally, strong constraints from the upper limits on the branching ratio of the tauonic $B_c$ decay, ${\rm BR}(B_c^{-} \to \tau^{-} \bar{\nu}_\tau) \lesssim 30 \%$ and $10\%$, imposed by the lifetime of $B_c$ meson~\cite{Alonso:2016oyd} and the LEP data taken at the $Z$ peak~\cite{Akeroyd:2017mhr}, respectively, have to be taken into account.
 
All these measurements on $b \to c\tau\bar{\nu}_\tau$ data point toward LFU violation and they are generally referred to as $b \to c\tau\bar{\nu}_\tau$ anomalies. Several model-independent studies of the effect of new physics (NP) operators regarding the most general dimension-six effective Lagrangian with the most recent $b \rightarrow c \tau \bar{\nu}_{\tau}$ data have been  explored~\cite{Iguro:2020keo,Iguro:2020cpg,Asadi:2019xrc,Murgui:2019czp,Mandal:2020htr,Cheung:2020sbq,Sahoo:2019hbu,Shi:2019gxi,
Bardhan:2019ljo,Blanke:2018yud,Blanke:2019qrx,Alok:2019uqc,Huang:2018nnq,Jung:2018lfu,Kumbhakar:2020jdz,
Angelescu:2018tyl,Feruglio:2018fxo,Iguro:2018vqb}. Sharing the same Lorentz structure as the SM, NP arising from left-handed vector operator $(\bar{c}\gamma_\mu P_L b)(\bar{\tau}\gamma^\mu P_L \nu_\tau)$ is still a preferred and feasible solution to address the anomalies, providing a good fit to the data~\cite{Iguro:2020keo,Iguro:2020cpg,Asadi:2019xrc,Murgui:2019czp,Mandal:2020htr,Cheung:2020sbq,Sahoo:2019hbu,Shi:2019gxi,
Bardhan:2019ljo,Blanke:2018yud,Blanke:2019qrx,Alok:2019uqc,Huang:2018nnq,Jung:2018lfu,Kumbhakar:2020jdz,
Angelescu:2018tyl,Feruglio:2018fxo,Iguro:2018vqb}. Different NP scenarios can be generated via this semi-tauonic operator. 
One interesting possibility to accommodate the anomalies consists in considering an extra left-handed gauge boson $W^\prime$~\cite{Bhattacharya:2014wla,Greljo:2015mma,Faroughy:2016osc,Abdullah:2018ets,Greljo:2018tzh,
Dasgupta:2018nzt, Boucenna:2016qad,Boucenna:2016wpr,Buttazzo:2017ixm,Bhattacharya:2016mcc,Kumar:2018kmr,
Gomez:2019xfw,Iguro:2018fni}. 
The opening works suggesting in the literature a $SU(2)_L$ triplet of massive vectors mostly coupled to the left-handed fermions of the third-generation (referred to as vector triplet model or vector boson model) were presented in Refs.~\cite{Bhattacharya:2014wla,Greljo:2015mma,Faroughy:2016osc}\footnote{Let us notice that a simultaneous explanation of both the $b \to c\tau\bar{\nu}_\tau$ and $b \to s\mu^+\mu^-$ anomalies have been also discussed within the vector boson model in Refs.~\cite{Bhattacharya:2014wla,Greljo:2015mma,Bhattacharya:2016mcc,Kumar:2018kmr}; however, this approach is beyond the scope of the present work.}. It was found that although the model can accommodate the $R(D^{(\ast)})$ anomalies, the framework is severely constrained by the direct  searches of neutral resonances decaying into $\tau^+ \tau^-$ pairs at ATLAS and CMS~\cite{Greljo:2015mma,Faroughy:2016osc}. Different $W^\prime$ boson scenarios (either UV completions and simplified models) have also been studied~\cite{Abdullah:2018ets,Greljo:2018tzh,
Dasgupta:2018nzt, Boucenna:2016qad,Boucenna:2016wpr,Buttazzo:2017ixm,Bhattacharya:2016mcc,Kumar:2018kmr,
Gomez:2019xfw,Iguro:2018fni} and complementary tests of these models with the searches for heavy $\tau\nu$ resonances performed at the LHC, showed an agreement with the constraints from ATLAS and CMS data (see, for instance, Refs.~\cite{Iguro:2018fni,Abdullah:2018ets,Greljo:2018tzh,Marzocca:2020ueu}). 

Additionally, alternative approaches regarding $W^\prime$ bosons associated with pure right-handed currents (involving a right-handed neutrino) have been discussed recently in the literature within different NP realizations~\cite{Greljo:2018tzh,Gomez:2019xfw,
Iguro:2018fni,Hayreter:2019dzc,He:2012zp,He:2017bft,Carena:2018cow,Babu:2018vrl,Asadi:2018wea,Greljo:2018ogz,Robinson:2018gza, Asadi:2018sym,Azatov:2018kzb,Cvetic:2017gkt}. Nevertheless, some recent analyses have shown that this right-handed neutrino interpretation seems to be disfavored by the LHC data~\cite{Shi:2019gxi,Greljo:2018tzh}.

On the other hand, LFU can also be tested through the ratio of leptonic decays of bottomonium meson $\Upsilon(nS)$~\cite{Aloni:2017eny}
\begin{equation} \label{R_Upsilon}
R_{\Upsilon(nS)} \equiv \frac{{\rm BR}(\Upsilon(nS) \to \tau^+\tau^-)}{{\rm BR}(\Upsilon(nS) \to \ell^+\ell^-)},
\end{equation}

\noindent with $n=1,2,3$ and $\ell=\mu,e$, providing a clean theoretical environment. Experimentally, the BABAR and CLEO Collaborations have reported the values~\cite{delAmoSanchez:2010bt,Besson:2006gj}
\begin{eqnarray}
R_{\Upsilon(1S)} &=&
\begin{cases}
\text{BABAR-10:} \ 1.005 \pm 0.013 \pm 0.022 \text{~\cite{delAmoSanchez:2010bt}}, \\
\text{SM:} \ 0.9924 \text{~\cite{Aloni:2017eny}},
\end{cases} \\
R_{\Upsilon(2S)} &=&
\begin{cases}
\text{CLEO-07:} \ 1.04 \pm 0.04 \pm 0.05 \text{~\cite{Besson:2006gj}}, \\
\text{SM:} \ 0.9940 \text{~\cite{Aloni:2017eny}},
\end{cases} \\
R_{\Upsilon(3S)} &=& 
\begin{cases}
\text{CLEO-07:} \ 1.05 \pm 0.08 \pm 0.05 \text{~\cite{Besson:2006gj}}, \\
\text{SM:} \ 0.9948 \text{~\cite{Aloni:2017eny}},
\end{cases} 
\end{eqnarray} 

\noindent where the theoretical uncertainty is typically of the order $\pm \mathcal{O}(10^{-5})$~\cite{Aloni:2017eny}. These measurements are in good accordance with the SM by $0.5\sigma, 0.8\sigma$, and $0.6\sigma$, respectively. Recently, in 2020 the BABAR experiment has released a new measurement on the ratio $R_{\Upsilon(3S)}$~\cite{Lees:2020kom}, whose value is
\begin{equation}
R_{\Upsilon(3S)}^{\rm BABAR-20} = 0.966 \pm 0.008 \pm 0.014,
\end{equation}

\noindent which improves the precision of the experimental value previously obtained by CLEO~\cite{Besson:2006gj}. Despite this improvement, the new value is below the SM expectation and shows an agreement at the $1.8\sigma$ level~\cite{Lees:2020kom}, in higher tension than CLEO. Moreover, averaging the CLEO-07~\cite{Besson:2006gj} and BABAR-20~\cite{Lees:2020kom} measurements we obtain 
 \begin{equation}
R_{\Upsilon(3S)}^{\rm Ave} = 0.968 \pm 0.016,
\end{equation}

\noindent which deviates at the $1.7\sigma$ level with respect to the SM prediction (uncertainties were taken in quadrature). Motivated by the tension generated by the new BABAR measurement on $R_{\Upsilon(3S)}$, it is intriguing to study its possible NP implications. As additional motivation, it is known that new physics scenarios (with left-handed neutrinos) aiming to provide an explanation to the $R(D^{(\ast)})$ anomalies also induce inevitable effects in the leptonic decay ratio $R_{\Upsilon(nS)}$~\cite{Aloni:2017eny}.
The connection between charged-current $b \to c \tau \bar{\nu}_{\tau}$ and neutral-current $b  \bar{b} \to \tau \bar{\tau}$ processes was first pointed out by the authors of Ref.~\cite{Faroughy:2016osc}, in which they performed a recast of existing $\tau^+\tau^-$ resonance searches at the CMS and ATLAS experiments, allowing one to set constraints on different simplified models addressing the $R(D^{(\ast)})$ anomalies. 

Keeping in mind the correlation between NP solutions to the charged-current $b \to c \tau \bar{\nu}_{\tau}$ anomalies and neutral-current $b  \bar{b} \to \tau \bar{\tau}$ processes~\cite{Faroughy:2016osc,Aloni:2017eny}, and to the light of the very recent BABAR result on $R_{\Upsilon(3S)}$~\cite{Lees:2020kom}, in this work we present a reanalysis of the extra gauge bosons within the vector triplet model that preferentially couples to third-generation fermions~\cite{Greljo:2015mma,Faroughy:2016osc}. 
A previous analysis addressing the $R(D^{(\ast)})$ anomalies and the complementary $R_{\Upsilon(nS)}$ in this model was presented in Ref.~\cite{Aloni:2017eny}, in which the authors found within $95\%$ confidence level the numerical values for the Wilson coefficients that minimize the observed anomaly in $R(D^{(\ast)})$, and the corresponding predictions for $R_{\Upsilon(nS)}$. This study was implemented by considering the 2016 HFLAV averages~\cite{Amhis:2016xyh}, which differ from the most recent 2019 HFLAV ones~\cite{Amhis:2019ckw,HFLAVsummer}. Here, by means of a different approach we carry out a robust phenomenological analysis of the parametric space of gauge couplings allowed by charged-current $b \to c \tau \bar{\nu}_\tau$ and $R_{\Upsilon(nS)}$ data. Particularly, for the $b \to c \tau \bar{\nu}_\tau$ data, we include  the polarizations of $D^*$ and the tau lepton associated with $\bar{B} \to D^\ast \tau \bar{\nu}_\tau$, the ratio $R(J/\psi)$, and the upper limit on BR$(B_c^{-} \to \tau^{-} \bar{\nu}_\tau)$, and we incorporate the forthcoming sensitivity of Belle II on $R(D^{(\ast)})$ measurements. In that sense, our work complements and extends the previous analysis performed in~\cite{Aloni:2017eny}. We will show that the vector triplet model is in conflict with the BABAR measurement of $R_{\Upsilon(3S)}$ and the $1\sigma$ range uncertainties cannot be explained in simultaneity with $b \to c \tau \bar{\nu}_\tau$ data.

The outline of this paper is organized as follows. In Sec.~\ref{model}, we briefly present the main features of the left-handed vector bosons model. A phenomenological analysis of the parametric space of gauge couplings allowed by charged-current and neutral-current data is presented in Sec.~\ref{analysis}. The main concluding remarks of this work are given in Sec.~\ref{Conclusion}.

\section{Left-handed vector bosons model} \label{model}
 
The SM is extended by including a color-neutral real $SU(2)_L$ triplet of massive vectors $W^\prime$ and $Z^\prime$ that coupled predominantly to left-handed (LH) fermions from the third-generation~\cite{Greljo:2015mma,Faroughy:2016osc}. The Lagrangian describing the interactions between fermions and vector boson is~\cite{Greljo:2015mma,Faroughy:2016osc} 
\begin{equation}
 \mathcal{L}^{\rm LH-VB} =  g_b \bar{Q}_{3}\frac{\sigma_a}{2} \gamma^\mu W^a_\mu Q_{3} + g_\tau \bar{L}_{3} \frac{\sigma_a}{2} \gamma^\mu W^a_\mu L_{3},
 \end{equation}

\noindent where $Q_{3}= (V_{cb}c_L, \ b_L)^T$ and $L_{3}= (\nu_{\tau L}, \ \tau_L)^T$ are the LH quark and lepton doublets, $\sigma_a$ ($a=1,2,3$) are the Pauli matrices, $V_{cb}$ is the associated Cabbibo-Kobayashi-Maskawa (CKM) matrix element, and $g_b$ and $g_\tau$ are the corresponding couplings of LH quarks and leptons to vector bosons, respectively. The down-type quark and charged-lepton mass eigenstate basis have been adopted for the LH fermion multiplets. After the heavy vector bosons are integrating out, the relevant charged-current $b\to c\tau\bar{\nu}_\tau$ and neutral-current $b \bar{b}\to \tau\bar{\tau}$ operators are given by~\cite{Greljo:2015mma}
\begin{eqnarray}
 \mathcal{L}_{\rm CC} &=& -\frac{g_b g_\tau}{2 M^2_{W^\prime}} V_{cb} (\bar{c} \gamma_\mu P_L b) (\bar{\tau} \gamma^\mu P_L \nu_{\tau}) + \ {\rm H.c.},  \\
  \mathcal{L}_{\rm NC} &=& -\frac{g_b g_\tau}{4 M^2_{Z^\prime}} (\bar{b} \gamma_\mu P_L b) (\bar{\tau} \gamma^\mu P_L \tau), 
\end{eqnarray}

\noindent respectively, where $M_{V}$ ($V=W^\prime,Z^\prime$) is the gauge boson mass. We are not assuming the existence of right-handed neutrinos within the model. Since the $W^\prime$ and $Z^\prime$ bosons couple primarily to the fermions from the third-generation, bounds coming from flavor-changing neutral-currents are avoided. According to electroweak precision data, it is required that gauge bosons are (almost) degenerate $M_{W^\prime} \simeq M_{Z^\prime}$~\cite{Faroughy:2016osc}. The NP effects are driven by the mass scale of the heavy mediators and the size of couplings to the third-generation of fermions $g_b$ and $g_\tau$. For simplicity, in further numerical analysis we will take these couplings to be real.

As it was mentioned above, an important caveat of the vector triplet model is that the parametric space required for the resolution of the $R(D^{(\ast)})$ anomalies and consistency with $\tau^+\tau^-$ resonance searches at the LHC (ATLAS and CMS) necessarily implies a very large $Z^\prime$ total decay width, $\Gamma_{Z^\prime} / M_{Z^\prime} = (g_\tau + 3g_b)/(48\pi) \gtrsim 30 \%$~\cite{Greljo:2015mma,Faroughy:2016osc}.  In Sec.~\ref{analysis} we will show that current $b \to c \tau \bar{\nu}_{\tau}$ data suggest that tension with constraints from ATLAS and CMS is now reduced.



\subsection{Contribution to the charged-current $b \to c \tau \bar{\nu}_{\tau}$ and neutral-current $b  \bar{b} \to \tau \bar{\tau}$ observables}

In the SM framework, the $b \to c \tau \bar{\nu}_{\tau}$ quark level processes are mediated by a virtual $W$ boson exchange. Within the NP scenarios discussed above, an extra $W^\prime$ boson leads to additional tree-level effective interactions, therefore, modifying the theoretical predictions for the observables associated with this charged-current transition. The ratios $R(M)$ ($M=D,D^\ast,J/\psi$), and the $D^\ast$ and $\tau$ longitudinal polarizations related with the channel $\bar{B} \to D^\ast \tau\bar{\nu}_\tau$ can be parametrized as~\cite{Gomez:2019xfw}
\begin{eqnarray}
R(M) &=& R(M)_{\rm SM} \big(\big|1 + C_{\rm VLL}^{bc\tau\nu_\tau} \big|^2\big),  \label{RD} \\
F_L(D^*) &=&  F_L(D^*)_{\rm SM} \ r_{D^\ast}^{-1}  \Big( \big|1 + C_{\rm VLL}^{bc\tau\nu_\tau}\big|^2 \big), \label{FLD} \\
P_\tau(D^*) &=&  P_\tau(D^*)_{\rm SM} \ r_{D^\ast}^{-1} \Big(|1 + C_{\rm VLL}^{bc\tau\nu_\tau}|^2 \big), \label{PTAU}
\end{eqnarray}

\noindent respectively, where $r_{D^\ast} = R(D^*) / R(D^*)_{\rm SM}$ and $C_{\rm VLL}^{bc\tau\nu_\tau}$ is the vector left-left (VLL) Wilson coefficient associated with the NP vector operators given by
\beq \label{CLL}
C_{\rm VLL}^{bc\tau\nu_\tau} = \dfrac{\sqrt{2}}{4G_F} \dfrac{g_{b}g_{\tau}}{M_{W^\prime}^2}, 
\eeq

\noindent with $G_F$ the Fermi coupling constant. Similarly, the tauonic decay $B_c^- \to \tau^- \bar{\nu}_{\tau}$ and the ratio $R(X_c)$ of inclusive semileptonic $B$ decays are also modified as~\cite{Gomez:2019xfw,Kamali:2018bdp}
\begin{eqnarray} \label{BRBc_modified}
{\rm BR}(B_c^- \to \tau^- \bar{\nu}_{\tau}) &=& {\rm BR}(B_c^- \to \tau^- \bar{\nu}_{\tau})_{\text{SM}}   \big(\big|1+C_{\rm VLL}^{bc\tau\nu_\tau} \big|^2 \big), \nonumber  \\  \\ 
R(X_c) &=& R(X_c)_{\rm SM} \Big( 1+ 1.147 \ \big|C_{\rm VLL}^{bc\tau\nu_\tau}\big|^2 \Big), \label{RXc}
\end{eqnarray}
\noindent respectively.

As concerns neutral-current process $b  \bar{b} \to \tau \bar{\tau}$, the leptonic decay ratio $R_{\Upsilon(nS)}$, Eq.\eqref{R_Upsilon}, is altered by the vector triplet model~\cite{Aloni:2017eny}. This ratio can be expressed as~\cite{Aloni:2017eny}
\begin{equation}
R_{\Upsilon(nS)}= \frac{(1-4x_\tau^2)^{1/2}}{\vert A_V^{\rm SM} \vert^2} \Big[ \vert A_V^{b\tau} \vert^2 (1+2x_\tau^2) + \vert B_V^{b\tau} \vert^2 (1- 4x_\tau^2) \Big],
\end{equation}

\noindent with $x_\tau = m_\tau/m_{\Upsilon(nS)}$, $\vert A_V^{\rm SM} \vert = - 4\pi\alpha Q_b$, and 
\begin{eqnarray}
A_V^{b\tau} &=& - 4\pi\alpha Q_b + \frac{m_{\Upsilon(nS)}^2}{4} C_{\rm VLL}^{bb\tau\tau}, \\
B_V^{b\tau} &=&- \frac{m_{\Upsilon(nS)}^2}{2} C_{\rm VLL}^{bb\tau\tau},
\end{eqnarray}

\noindent where 
\begin{equation}
C_{\rm VLL}^{bb\tau\tau} = \frac{g_b g_\tau}{4M_{W^\prime}^2}.
\end{equation}

\noindent It is straightforward to see the relation between charged and neutral coefficients, $C_{\rm VLL}^{bc\tau\nu_\tau}  = (\sqrt{2}/G_F) C_{\rm VLL}^{bb\tau\tau}$. In the next section, we will present a phenomenological analysis of the parametric space of gauge couplings allowed by $b \to c \tau \bar{\nu}_{\tau}$ and $b  \bar{b} \to \tau \bar{\tau}$ data.

\section{Phenomenological study} \label{analysis}

\begin{table*}[!t]
\centering
\renewcommand{\arraystretch}{1.2}
\renewcommand{\arrayrulewidth}{0.8pt}
\begin{tabular}{ccccc}
\hline \hline                                                                                                                           
Dataset & ($g_b,g_\tau$) & $\chi^2_{\rm min}/N_{\rm dof}$ & $p$-value (\%) & $\rm{pull_{SM}}$ \\                                                                                                                             
\hline
$b \to c \tau \bar{\nu}_\tau$ & (2.99,1.54) & 1.04 & 39.0 & 3.72  \\
$b \to c \tau \bar{\nu}_\tau$ + $R_\Upsilon$ old & (3.05,1.52) & 0.79 & 61.3 & 3.75  \\
$b \to c \tau \bar{\nu}_\tau$ + $R_\Upsilon$ with BABAR-20 & (3.27,1.39) & 1.20 & 29.3 & 3.68  \\
$b \to c \tau \bar{\nu}_\tau$ + $R_\Upsilon$ combined & (3.05,1.52) & 1.11 & 35.3 & 3.68  \\
\hline\hline
\end{tabular}
\caption{BFP values of gauge couplings, $\chi^2_{\rm min}/N_{\rm dof}$, $p$-value, and $\rm{pull_{SM}}$ for different datasets of observables.} \label{fit}
\end{table*} 

To provide a robust phenomenological study we consider all of the charged-current transition $b \to c \tau \bar{\nu}_\tau$ observables, namely, the ratios $R(D^{(\ast)})$ (HFLAV 2019 averages), $R(J/\psi)$, $R(X_c)$; the polarizations $P_\tau(D^\ast), F_L(D^\ast)$; and the upper limit BR$(B_c^{-} \to \tau^{-} \bar{\nu}_\tau) < 10 \%$. We will refer to this set as the  $b \to c \tau \bar{\nu}_\tau$ data. On the other hand, regarding the neutral-current observables $R_{\Upsilon(nS)}$, we will take into account three different datasets:
\begin{enumerate}
\item $R_{\Upsilon}$ old data: $R_{\Upsilon(1S)}$ BABAR-10~\cite{delAmoSanchez:2010bt}, $R_{\Upsilon(2S)}$ CLEO-07~\cite{Besson:2006gj}, and $R_{\Upsilon(3S)}$ CLEO-07~\cite{Besson:2006gj}.
\item $R_{\Upsilon}$ with BABAR-20 data: $R_{\Upsilon(1S)}$ BABAR-10~\cite{delAmoSanchez:2010bt}, $R_{\Upsilon(2S)}$ CLEO-07~\cite{Besson:2006gj}, and $R_{\Upsilon(3S)}$ BABAR-20~\cite{Lees:2020kom}.
\item $R_{\Upsilon}$ combined data: $R_{\Upsilon(1S)}$ BABAR-10~\cite{delAmoSanchez:2010bt}, $R_{\Upsilon(2S)}$ CLEO-07~\cite{Besson:2006gj}, and $R_{\Upsilon(3S)}$ average of CLEO-07~\cite{Besson:2006gj} and BABAR-20~\cite{Lees:2020kom},
\end{enumerate}

\noindent with $\Upsilon \equiv \Upsilon(nS)$ for simplicity. The purpose of these sets is to estimate the impact of the very recent BABAR measurement on $R_{\Upsilon(3S)}$~\cite{Lees:2020kom}. Furthermore, we complement this analysis by exploring two plausible scenarios on the $R(D^{(\ast)})$ future measurements in the ongoing Belle II experiment~\cite{Kou:2018nap}. The two projected scenarios are as follows~\cite{Cardozo:2020uol}, Belle II-P1: Belle II measurements on $R(D^{(\ast)})$ keep the central values of Belle combination averages with the projected Belle II sensitivities for $50 \ {\rm ab}^{-1}$~\cite{Kou:2018nap}; and Belle II-P2: Belle II measurements on $R(D^{(\ast)})$ are in agreement with the current SM predictions at the $0.1\sigma$ level with the projected Belle II sensitivities for $50 \ {\rm ab}^{-1}$~\cite{Kou:2018nap}. These Belle II future implications on a $W^{\prime}$ boson scenario have not been explored so far in previous works. 

Bearing in mind the above-mentioned observables, we perform a standard $\chi^2 \equiv \chi^2(g_b,g_\tau)$ function analysis in order to prove whether it is possible to adjust the deviations of the SM predictions in the simplified extra gauge bosons model described in Sec.~\ref{model}. We consider the experimental correlation value $-0.38$ between $R(D)$ and $R(D^\ast)$ from HFLAV~\cite{Amhis:2019ckw,HFLAVsummer}. We determine the regions in the parameter space favored by the experimental data.

\begin{figure*}[!t]
\centering
\includegraphics[scale=0.27]{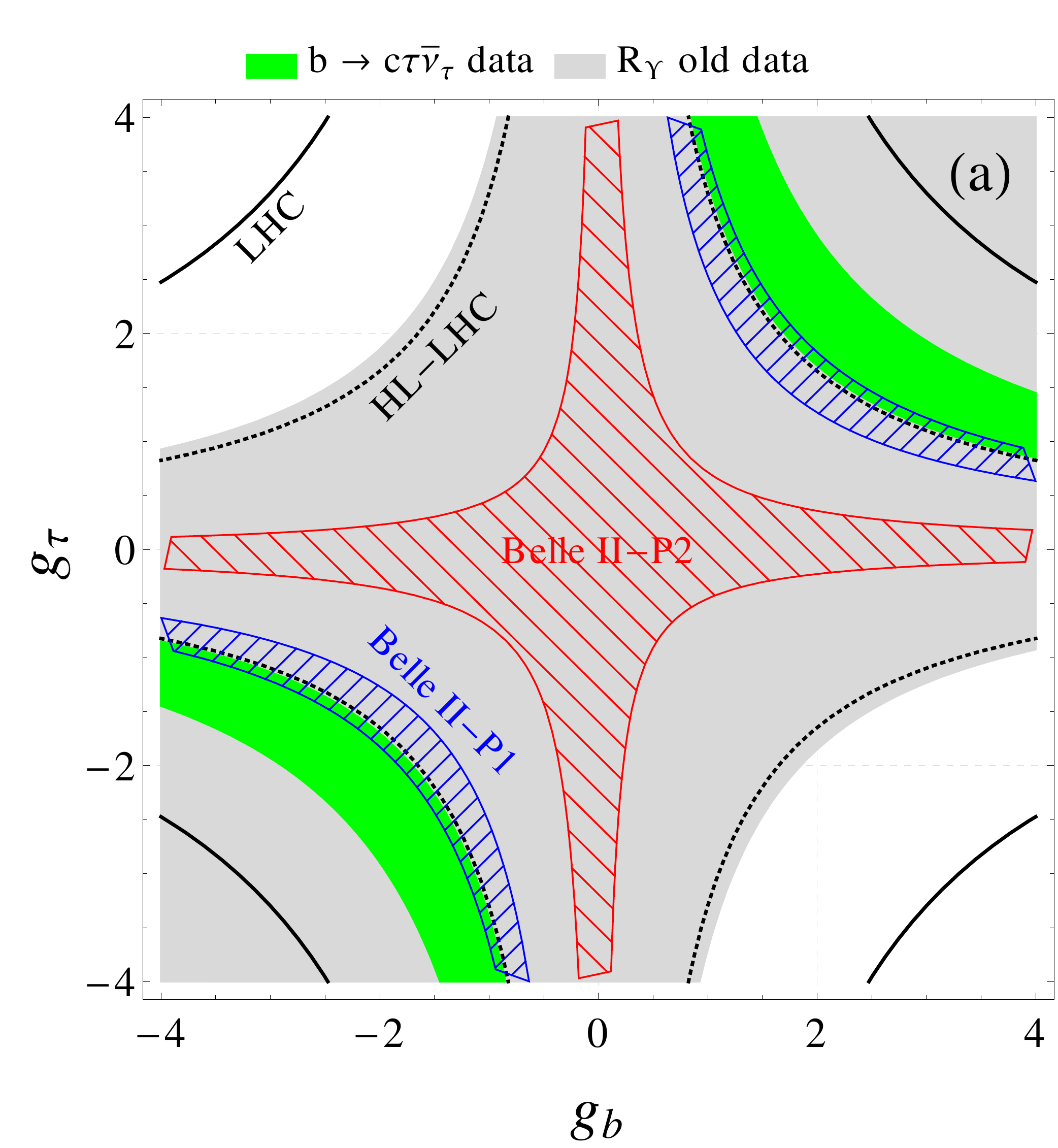} \ 
\includegraphics[scale=0.27]{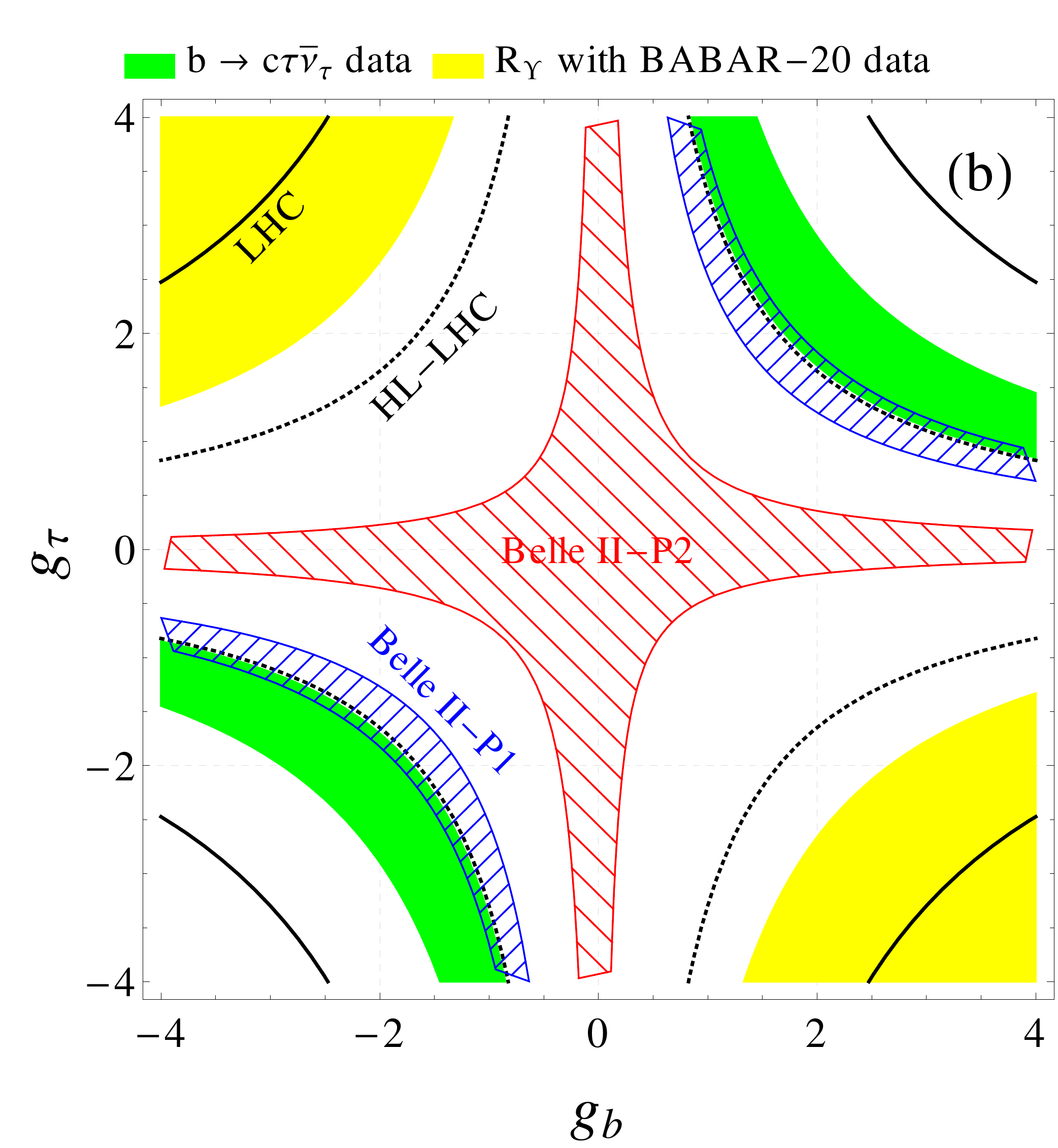}\ 
\includegraphics[scale=0.27]{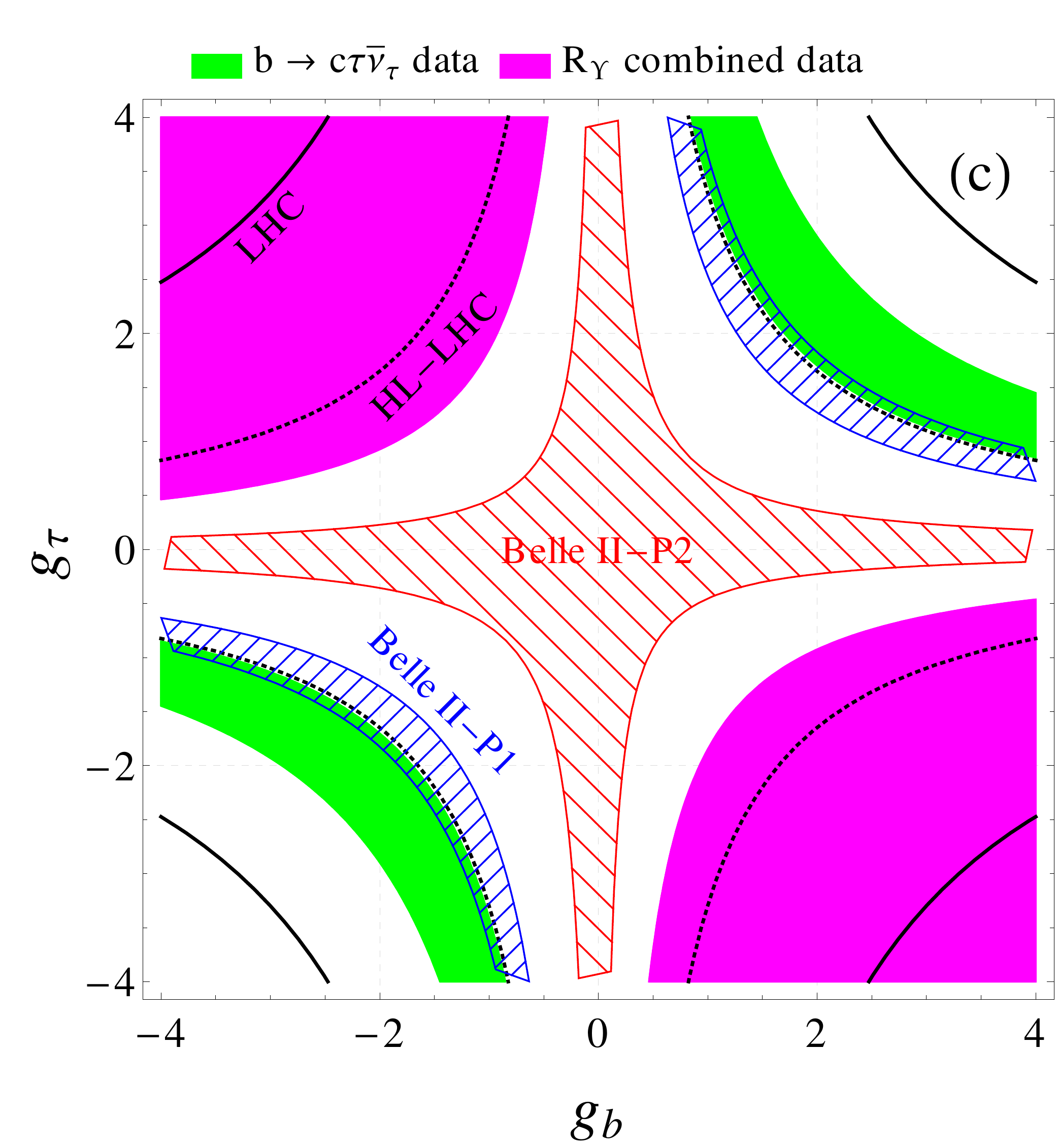}
\caption{\small The $1\sigma$ allowed parameter space in the ($g_b,g_\tau$) plane for the current $b\to c \tau \bar{\nu}_{\tau}$ data [green region] and (a) $R_\Upsilon$ old data [gray region], (b) $R_\Upsilon$ with BABAR-20 data [yellow region], and (c) $R_\Upsilon$ combined data [magenta region], for $M_{W^\prime} = 1 \ {\rm TeV}$. The projection Belle II-P1 (Belle II-P2) for an integrated luminosity of $50 \ {\rm ab}^{-1}$ is represented by the blue (red) hatched region. The inner black contour lines illustrate the permitted regions from LHC bounds (solid line) and HL-LHC prospects (dotted line).}
\label{ParameterSpace1}
\end{figure*}

\subsection{Parametric space ($g_b,g_{\tau}$)}

After fitting different sets of observables, we display in Table~\ref{fit} our results of the best-fit point (BFP) values on the gauge couplings  ($g_b,g_{\tau}$), the ratio of the minimum of the $\chi^2$ function and number of degrees of freedom ($\chi^2_{\rm min}/N_{\rm dof}$), the $p$-value, and the pull of the SM $\rm{pull_{SM}}=\sqrt{\chi^2_{\rm SM} - \chi^2_{\rm min}}$, with $\chi^2_{\rm SM} = \chi^2(0)$. In order to keep the couplings in the perturbative regime ($\sim \sqrt{4\pi}$), we took a benchmark $W^{\prime}$ mass value of $M_{W^\prime}= 1$~TeV in our analysis.  There is no tension with the current LHC constraints for the $M_{W^\prime}$ (which are above 4~TeV) since we are assuming zero couplings to the first and second families. For a $W^\prime$ dominantly coupled to a third family, a $W^{\prime}$ mass value of $\sim 1$ TeV is compatible with LHC bounds (see, for instance, Refs.~\cite{Iguro:2020keo,Marzocca:2020ueu,Hayreter:2019dzc}.) From Table~\ref{fit}, it is observed that with only $b\to c \tau \bar{\nu}_{\tau}$ data is a good fit obtained, as expected, with a $p$-value $= 39\%$. When $b \to c \tau \bar{\nu}_\tau$ and $R_\Upsilon$ old data are joined together, a better fit is obtained with a larger $p$-value of $61.3 \%$. This indicates that the extra gauge bosons model can simultaneously explain both charged-current and neutral-current data of $b$-flavored mesons. However, once the BABAR measurement on $R_{\Upsilon(3S)}$~\cite{Lees:2020kom} is incorporated into the fit, through either $R_\Upsilon$ with BABAR-20 or $R_\Upsilon$ combined data, it induces tension in the analysis, causing the quality of the fit to decrease (smaller $p$-value), but maintaining almost the same value of BFP and $\rm{pull_{SM}}$. In turn, BABAR's result~\cite{Lees:2020kom} seems to challenge this NP explanation.

In Figs.~\ref{ParameterSpace1}(a),~\ref{ParameterSpace1}(b), and~\ref{ParameterSpace1}(c) we show the $1\sigma$ allowed parameter space in the ($g_b,g_{\tau}$) plane, where the gray, yellow, and magenta regions are obtained by considering $R_\Upsilon$ old data, $R_\Upsilon$ with BABAR-20 data, and $R_\Upsilon$ combined data, respectively. In all of the panels, the green region represents the allowed region by the charged-current transition $b\to c \tau \bar{\nu}_{\tau}$ data, and the projection Belle II-P1 (Belle II-P2) for an integrated luminosity of $50 \ {\rm ab}^{-1}$ is represented by the blue (red) hatched region. To further extend our analysis, the inner black contour lines illustrate the permitted regions from LHC bounds (solid line) and the prospects at the high-luminosity (HL)-LHC (dotted line)~\cite{Iguro:2020keo,Marzocca:2020ueu}.~\footnote{These contours have been obtained by taking into account the LHC bounds on the left-handed vector WC of $\vert  C_{\rm VLL}^{cb\tau\nu_\tau} \vert \simeq 0.3$ and the future prospects values at HL-LHC of $\vert  C_{\rm VLL}^{cb\tau\nu_\tau} \vert \simeq 0.1$, evaluated at 1 TeV scale~\cite{Iguro:2020keo,Marzocca:2020ueu}.}
From Fig.~\ref{ParameterSpace1}(a) one can note that it is possible to get an allowed region on the parameter space to account for a joint explanation to the $b \to c \tau \bar{\nu}_\tau$ and $R_\Upsilon$ old data, in consistency with LHC and HL-LHC bounds. As for Figs.~\ref{ParameterSpace1}(b) and~\ref{ParameterSpace1}(c), the datasets $R_\Upsilon$ with BABAR-20 and $R_\Upsilon$ combined prove a different parametric space not compatible with $b\to c \tau \bar{\nu}_{\tau}$ data. Thus, we confirm that the recent BABAR results on $R_{\Upsilon(3S)}$ generates tension; therefore, charged-current and neutral-current data of $b$-flavored mesons cannot be addressed simultaneously in this model. Only relaxing the $R_{\Upsilon(3S)}$ experimental uncertainties to the $2\sigma$ level can a common allowed region be obtained. Regarding the Belle II experiment, the projection Belle II-P2 indicates that the parametric space would be severely constrained, but still allow a window for significant NP contributions. Remarkably, the Belle II-P2 scenario would provide stronger bounds on the ($g_b,g_{\tau}$) plane than prospects at the HL-LHC. 

On the other hand, as concerns the vector triplet model interpretation to the most recent $b \to c \tau \bar{\nu}_\tau$ data and its consistency with LHC searches for $Z^\prime$ resonances decaying to $\tau^+\tau^-$~\cite{Greljo:2015mma,Faroughy:2016osc}, it is observed that using the BFP gauge couplings values we get $\Gamma_{Z^\prime} / M_{Z^\prime} \simeq 20 \%$ for $M_{Z^\prime} = 1$ TeV, implying a decrease in the tension. In addition, future Belle II sensitivity on $R(D^{(\ast)})$ would point to smaller widths $\Gamma_{Z^\prime} / M_{Z^\prime} \simeq (1 - 5) \%$ within the allowed regions by LHC bounds~\cite{Greljo:2015mma,Faroughy:2016osc}.

In summary, the recent BABAR results on $R_{\Upsilon(3S)}$ hint toward a new anomalous measurement. Here, we exemplified its implications on the vector triplet model. Because in a typical electroweak extension of the SM the charged current parameters and the corresponding ones in the neutral sector are related, this analysis can be carried out in most of the models involving the structure $(\bar{c}\gamma_\mu P_L b)(\bar{\tau}\gamma^\mu P_L \nu_\tau)$. Thus, our conclusions can be extrapolated to those models. Alternative NP scenarios can give rise to the same left-left vector operator such as vector leptoquark models; therefore, it is interesting to explore the possible effects of $R_{\Upsilon(3S)}$ on these scenarios~\cite{Garcia:2021}.





\section{Concluding remarks} \label{Conclusion}

New physics scenarios aiming to provide an explanation to the LFU violation anomalies reported in the charged-current observables of semileptonic $B$ meson decays also induce effects in the neutral-current observables of bottomonium mesons $R_{\Upsilon(nS)}$, with $n=1,2,3$. Motivated by the very recent BABAR measurement on $R_{\Upsilon(3S)}$, we revisited the simplified scenario of extra massive gauge bosons ($W^\prime$ and $Z^\prime$) that coupled predominantly to leptons of the third generation (involving LH neutrinos), proposed as a viable solution to the $b \to c \tau \bar{\nu}_\tau$ anomalies. We performed a robust phenomenological analysis of the parametric space of gauge couplings allowed by the most recent charged-current $b \to c \tau \bar{\nu}_\tau$ and neutral-current $b \bar{b} \to \tau\bar{\tau}$ data. As the main result of our analysis, it is found that the BABAR measurement of $R_{\Upsilon(3S)}$ is particularly challenging and the $1\sigma$ range uncertainties cannot be explained simultaneously with charged-current $b \to c \tau \bar{\nu}_\tau$ data within the LH vector bosons model. Therefore, this NP scenario seems to be disfavored by BABAR data. In order to clarify this situation, future $R_{\Upsilon(3S)}$ measurements in the ongoing experiments Belle II and LHCb will be a matter of importance to confirm or refute the discrepancy.


\acknowledgments
J. H. Mu\~{n}oz is grateful to Oficina de Investigaciones of Universidad del Tolima for financial support of Project No. 290130517. The work of N. Quintero has been financially supported by MINCIENCIAS and Universidad del Tolima through Convocatoria Estancias Postdoctorales No. 848-2019 (Contract No. 834-2020). E. Rojas acknowledges financial support from the “Vicerrectoría de Investigaciones e Interacción Social VIIS de la Universidad de Nariño”, Projects No. 1928 and No. 2172.

\FloatBarrier


\begin{thebibliography}{99}


\bibitem{Lees:2012xj} 
  J.~P.~Lees {\it et al.} [BaBar Collaboration],
  Evidence for an excess of $\bar{B} \to D^{(*)} \tau^-\bar{\nu}_\tau$ decays,
  Phys.\ Rev.\ Lett.\  {\bf 109}, 101802 (2012)
  [arXiv:1205.5442 [hep-ex]].
  
\bibitem{Lees:2013uzd} 
  J.~P.~Lees {\it et al.} [BaBar Collaboration],
  Measurement of an Excess of $\bar{B} \to D^{(*)}\tau^- \bar{\nu}_\tau$ Decays and Implications for Charged Higgs Bosons,
  Phys.\ Rev.\ D {\bf 88}, no. 7, 072012 (2013)
  [arXiv:1303.0571 [hep-ex]].


\bibitem{Huschle:2015rga} 
  M.~Huschle {\it et al.} [Belle Collaboration],
  Measurement of the branching ratio of $\bar{B} \to D^{(\ast)} \tau^- \bar{\nu}_\tau$ relative to $\bar{B} \to D^{(\ast)} \ell^- \bar{\nu}_\ell$ decays with hadronic tagging at Belle,
  Phys.\ Rev.\ D {\bf 92}, no. 7, 072014 (2015)
  [arXiv:1507.03233 [hep-ex]].


\bibitem{Sato:2016svk} 
  Y.~Sato {\it et al.} [Belle Collaboration],
  Phys.\ Rev.\ D {\bf 94}, no. 7, 072007 (2016)
  [arXiv:1607.07923 [hep-ex]].
  
\bibitem{Hirose:2017vbz} 
  S.~Hirose [Belle Collaboration],
  $\bar{B} \rightarrow D^{(*)} \tau^- \bar{\nu}_\tau$ and Related Tauonic Topics at Belle,
  arXiv:1705.05100 [hep-ex].
  
\bibitem{Aaij:2015yra} 
  R.~Aaij {\it et al.} [LHCb Collaboration],
  Measurement of the ratio of branching fractions $\mathcal{B}(\bar{B}^0 \to D^{*+}\tau^{-}\bar{\nu}_{\tau})/\mathcal{B}(\bar{B}^0 \to D^{*+}\mu^{-}\bar{\nu}_{\mu})$,
  Phys.\ Rev.\ Lett.\  {\bf 115}, no. 11, 111803 (2015)
  Erratum: [Phys.\ Rev.\ Lett.\  {\bf 115}, no. 15, 159901 (2015)]
  [arXiv:1506.08614 [hep-ex]].


\bibitem{Aaij:2017deq} 
  R.~Aaij {\it et al.} [LHCb Collaboration],
  Test of Lepton Flavor Universality by the measurement of the $B^0 \to D^{*-} \tau^+ \nu_{\tau}$ branching fraction using three-prong $\tau$ decays,
  Phys.\ Rev.\ D {\bf 97}, no. 7, 072013 (2018)
  [arXiv:1711.02505 [hep-ex]].


\bibitem{Aaij:2017uff} 
  R.~Aaij {\it et al.} [LHCb Collaboration],
  Measurement of the ratio of the $B^0 \to D^{*-} \tau^+ \nu_{\tau}$ and $B^0 \to D^{*-} \mu^+ \nu_{\mu}$ branching fractions using three-prong $\tau$-lepton decays,
  Phys.\ Rev.\ Lett.\  {\bf 120}, no. 17, 171802 (2018)
  [arXiv:1708.08856 [hep-ex]].

\bibitem{Belle:2019rba}
G.~Caria \textit{et al.} [Belle Collaboration], Measurement of $\mathcal{R}(D)$ and $\mathcal{R}(D^*)$ with a Semileptonic Tagging Method, 
Phys. Rev. Lett. \textbf{124} (2020) no.16, 161803
[arXiv:1910.05864 [hep-ex]].

\bibitem{Hirose:2017dxl} 
  S.~Hirose {\it et al.} [Belle Collaboration],
  Measurement of the $\tau$ lepton polarization and $R(D^*)$ in the decay $\bar{B} \rightarrow D^* \tau^- \bar{\nu}_\tau$ with one-prong hadronic $\tau$ decays at Belle,
  Phys.\ Rev.\ D {\bf 97}, no. 1, 012004 (2018)
  [arXiv:1709.00129 [hep-ex]].


\bibitem{Hirose:2016wfn} 
  S.~Hirose {\it et al.} [Belle Collaboration],
  Measurement of the $\tau$ lepton polarization and $R(D^*)$ in the decay $\bar{B} \to D^* \tau^- \bar{\nu}_\tau$,
  Phys.\ Rev.\ Lett.\  {\bf 118}, no. 21, 211801 (2017)
  [arXiv:1612.00529 [hep-ex]].
  
\bibitem{Amhis:2019ckw} 
  Y.~S.~Amhis {\it et al.} [HFLAV Collaboration],
  Averages of $b$-hadron, $c$-hadron, and $\tau$-lepton properties as of 2018,
  arXiv:1909.12524 [hep-ex].
  
\bibitem{HFLAVsummer}
For updated results see HFLAV average of $R(D^{(\ast)})$ for Spring 2019 in \url{https://hflav-eos.web.cern.ch/hflav-eos/semi/spring19/html/RDsDsstar/RDRDs.html}.


\bibitem{Aaij:2017tyk} 
R.~Aaij {\it et al.} (LHCb Collaboration), Measurement of the ratio of branching fractions $\mathcal{B}(B_c^+\,\to\,J/\psi\tau^+\nu_\tau)$/$\mathcal{B}(B_c^+\,\to\,J/\psi\mu^+\nu_\mu)$,  Phys. Rev. Lett. \textbf{120}, 121801 (2018) \href{http://arxiv.org/abs/1711.05623}{[arXiv:1711.05623 [hep-ex]]}.

\bibitem{Harrison:2020nrv}
J.~Harrison \textit{et al.} [LATTICE-HPQCD], $R(J/\psi)$ and $B_c^- \rightarrow J/\psi \ell^-\bar{\nu}_\ell$ Lepton Flavor Universality Violating Observables from Lattice QCD,
Phys. Rev. Lett. \textbf{125}, no.22, 222003 (2020)
[arXiv:2007.06956 [hep-lat]].

\bibitem{Harrison:2020gvo}
J.~Harrison \textit{et al.} [HPQCD], $B_c \rightarrow J/\psi$ form factors for the full $q^2$ range from lattice QCD,
Phys. Rev. D \textbf{102}, no.9, 094518 (2020)
[arXiv:2007.06957 [hep-lat]].

\bibitem{Abdesselam:2019wbt} 
  A.~Abdesselam {\it et al.} [Belle Collaboration],
  Measurement of the $D^{\ast-}$ polarization in the decay $B^0 \to D^{\ast -}\tau^+\nu_{\tau}$,
  arXiv:1903.03102 [hep-ex].

\bibitem{Tanaka:2012nw} 
M.~Tanaka and R.~Watanabe, New physics in the weak interaction of $\bar B\to D^{(*)}\tau\bar\nu$, Phys.\ Rev.\ D {\bf 87}, 034028 (2013)
\href{http://arxiv.org/abs/1212.1878}{[arXiv:1212.1878 [hep-ph]]}.

\bibitem{Alok:2016qyh} 
A.~K.~Alok, D.~Kumar, S.~Kumbhakar and S.~U.~Sankar, $D^{*}$ polarization as a probe to discriminate new physics in $\bar{B}\to D^{*} \tau \bar{\nu}$, Phys.\ Rev.\ D {\bf 95}, 115038 (2017)
\href{http://arxiv.org/abs/1606.03164}{[arXiv:1606.03164 [hep-ph]]}.

\bibitem{Alonso:2016oyd} 
R.~Alonso, B.~Grinstein and J.~Martin Camalich, Lifetime of $B_c^-$ Constrains Explanations for Anomalies in  $B\to D^{(*)}\tau\nu$, Phys.\ Rev.\ Lett.\  {\bf 118}, 081802 (2017). \href{http://arxiv.org/abs/1611.06676}{[arXiv:1611.06676 [hep-ph]]}   

\bibitem{Akeroyd:2017mhr}
A.~G.~Akeroyd and C.~H.~Chen, Constraint on the branching ratio of $B_c \to \tau \nu$ from LEP1 and consequences for R(D(*)) anomaly, Phys. Rev. D \textbf{96}, 075011 (2017). \href{http://arxiv.org/abs/1708.04072}{[arXiv:1708.04072 [hep-ph]]}.


\bibitem{Iguro:2020keo}
S.~Iguro, M.~Takeuchi and R.~Watanabe, Testing Leptoquark/EFT in $\bar B \to D^{(*)} l\bar\nu$ at the LHC,[arXiv:2011.02486 [hep-ph]].

\bibitem{Iguro:2020cpg}
S.~Iguro and R.~Watanabe, Bayesian fit analysis to full distribution data of $ \overline{\mathrm{B}}\to {\mathrm{D}}^{\left(\ast \right)}\mathrm{\ell}\overline{\nu }:\left|{\mathrm{V}}_{\mathrm{cb}}\right| $ determination and new physics constraints, JHEP \textbf{08}, no.08, 006 (2020)
[arXiv:2004.10208 [hep-ph]].

\bibitem{Asadi:2019xrc} 
  P.~Asadi and D.~Shih,
  Maximizing the Impact of New Physics in $b\rightarrow c \tau \nu$ Anomalies, Phys.\ Rev.\ D {\bf 100}, no. 11, 115013 (2019)
  [arXiv:1905.03311 [hep-ph]].

\bibitem{Murgui:2019czp} 
  C.~Murgui, A.~Peñuelas, M.~Jung and A.~Pich, Global fit to $b \to c \tau \nu$ transitions, JHEP {\bf 1909}, 103 (2019)
  [arXiv:1904.09311 [hep-ph]].

\bibitem{Mandal:2020htr}
R.~Mandal, C.~Murgui, A.~Peñuelas and A.~Pich, The role of right-handed neutrinos in $b \to c \tau \bar{\nu}$ anomalies, JHEP \textbf{08}, 022 (2020)
[arXiv:2004.06726 [hep-ph]].

\bibitem{Cheung:2020sbq} 
  K.~Cheung, Z.~R.~Huang, H.~D.~Li, C.~D.~Lü, Y.~N.~Mao and R.~Y.~Tang,
  Revisit to the $b\to c\tau\nu$ transition: in and beyond the SM,
  arXiv:2002.07272 [hep-ph].
  
\bibitem{Sahoo:2019hbu} 
  S.~Sahoo and R.~Mohanta, Investigating the role of new physics in $b \to c \tau \bar \nu_\tau$ transitions,
  arXiv:1910.09269 [hep-ph].

\bibitem{Shi:2019gxi} 
  R.~X.~Shi, L.~S.~Geng, B.~Grinstein, S.~Jäger and J.~Martin Camalich, Revisiting the new-physics interpretation of the $b\to c\tau\nu$ data, JHEP {\bf 1912}, 065 (2019)
  [arXiv:1905.08498 [hep-ph]].
  
  \bibitem{Bardhan:2019ljo}
  D.~Bardhan and D.~Ghosh, $B$ -meson charged current anomalies: The post-Moriond 2019 status,
  Phys.\ Rev.\ D {\bf 100} (2019) no.1,  011701
  [arXiv:1904.10432 [hep-ph]].
 
 
\bibitem{Blanke:2018yud} 
M.~Blanke, A.~Crivellin, S.~de Boer, T.~Kitahara, M.~Moscati, U.~Nierste and I.~Nisandzic, Impact of polarization observables and $ B_c\to \tau \nu$ on new physics explanations of the $b\to c \tau \nu$ anomaly,'
Phys. Rev. D \textbf{99}, no.7, 075006 (2019)
[arXiv:1811.09603 [hep-ph]].

\bibitem{Blanke:2019qrx}
M.~Blanke, A.~Crivellin, T.~Kitahara, M.~Moscati, U.~Nierste and I.~Nisandzic, Addendum to “Impact of polarization observables and $B_c\to \tau \nu$ on new physics explanations of the $b\to c \tau \nu$ anomaly",
Phys. Rev. D \textbf{100}, 035035 (2019)
[arXiv:1905.08253 [hep-ph]].

\bibitem{Alok:2019uqc}
A.~K.~Alok, D.~Kumar, S.~Kumbhakar and S.~Uma Sankar, Solutions to $R_D$-$R_{D^*}$ in light of Belle 2019 data,
Nucl. Phys. B \textbf{953}, 114957 (2020)
[arXiv:1903.10486 [hep-ph]].

\bibitem{Huang:2018nnq}
Z.~R.~Huang, Y.~Li, C.~D.~Lu, M.~A.~Paracha and C.~Wang, Footprints of New Physics in $b\to c\tau\nu$ Transitions, Phys. Rev. D \textbf{98}, no.9, 095018 (2018)
[arXiv:1808.03565 [hep-ph]].

\bibitem{Jung:2018lfu} 
M.~Jung and D.~M.~Straub, Constraining new physics in $b\to c\ell\nu$ transitions,  JHEP {\bf 1901}, 009 (2019)
[arXiv:1801.01112 [hep-ph]].
  
\bibitem{Kumbhakar:2020jdz}
S.~Kumbhakar, Signatures of complex new physics in $b\to c\tau\bar{\nu}$ transitions,
Nucl. Phys. B \textbf{963}, 115297 (2021)
[arXiv:2007.08132 [hep-ph]].

 
\bibitem{Angelescu:2018tyl}
A.~Angelescu, D.~Be\v{c}irevi\'c, D.~A.~Faroughy and O.~Sumensari, Closing the window on single leptoquark solutions to the $B$-physics anomalies, JHEP \textbf{10}, 183 (2018)
[arXiv:1808.08179 [hep-ph]].

\bibitem{Feruglio:2018fxo}
F.~Feruglio, P.~Paradisi and O.~Sumensari, Implications of scalar and tensor explanations of $R_{D^{(\ast)}}$,
JHEP \textbf{11}, 191 (2018)
[arXiv:1806.10155 [hep-ph]].

\bibitem{Iguro:2018vqb}
S.~Iguro, T.~Kitahara, Y.~Omura, R.~Watanabe and K.~Yamamoto, D$^{*}$ polarization vs. $ {R}_{D^{\left(\ast \right)}} $ anomalies in the leptoquark models, JHEP \textbf{02}, 194 (2019)
[arXiv:1811.08899 [hep-ph]].



\bibitem{Bhattacharya:2014wla}
B.~Bhattacharya, A.~Datta, D.~London and S.~Shivashankara, Simultaneous Explanation of the $R_K$ and $R(D^{(*)})$ Puzzles, Phys. Lett. B \textbf{742}, 370-374 (2015)
[arXiv:1412.7164 [hep-ph]].

\bibitem{Greljo:2015mma} 
A.~Greljo, G.~Isidori and D.~Marzocca, On the breaking of Lepton Flavor Universality in B decays,  JHEP {\bf 1507}, 142 (2015)   [arXiv:1506.01705 [hep-ph]].

\bibitem{Faroughy:2016osc} 
D.~A.~Faroughy, A.~Greljo and J.~F.~Kamenik, Confronting lepton flavor universality violation in B decays with high-$p_T$ tau lepton searches at LHC, Phys.\ Lett.\ B {\bf 764}, 126 (2017). \href{http://arxiv.org/abs/1609.07138}{[arXiv:1609.07138 [hep-ph]]} 

\bibitem{Abdullah:2018ets} 
M.~Abdullah, J.~Calle, B.~Dutta, A.~Fl\'{o}rez and D.~Restrepo, Probing a simplified $W^{\prime}$ model of $R(D^{(\ast)})$ anomalies using $b$-tags, $\tau$ leptons and missing energy, Phys.\ Rev.\ D {\bf 98}, 055016 (2018) [arXiv:1805.01869 [hep-ph]].

\bibitem{Greljo:2018tzh} 
 A.~Greljo, J.~Martin Camalich and J.~D.~Ruiz-\'{A}lvarez, Mono-$\tau$ Signatures at the LHC Constrain Explanations of B-decay Anomalies, Phys.\ Rev.\ Lett.\  {\bf 122}, 131803 (2019) [arXiv:1811.07920 [hep-ph]].

\bibitem{Dasgupta:2018nzt} 
  S.~Dasgupta, U.~K.~Dey, T.~Jha and T.~S.~Ray, Status of a flavor-maximal nonminimal universal extra dimension model, Phys.\ Rev.\ D {\bf 98}, 055006 (2018)   [arXiv:1801.09722 [hep-ph]].

\bibitem{Boucenna:2016qad} 
S.~M.~Boucenna, A.~Celis, J.~Fuentes-Martin, A.~Vicente and J.~Virto, Phenomenology of an $SU(2) \times SU(2) \times U(1)$ model with lepton-flavour non-universality, JHEP {\bf 1612}, 059 (2016)  \href{http://arxiv.org/abs/1608.01349}{[arXiv:1608.01349 [hep-ph]]}.

\bibitem{Boucenna:2016wpr}
S.~M.~Boucenna, A.~Celis, J.~Fuentes-Martin, A.~Vicente and J.~Virto, Non-abelian gauge extensions for B-decay anomalies, Phys.\ Lett.\ B {\bf 760}, 214 (2016) \href{http://arxiv.org/abs/1604.03088}{[arXiv:1604.03088 [hep-ph]]}.

\bibitem{Buttazzo:2017ixm}
D.~Buttazzo, A.~Greljo, G.~Isidori and D.~Marzocca, B-physics anomalies: a guide to combined explanations,
JHEP \textbf{11}, 044 (2017)
[arXiv:1706.07808 [hep-ph]].

\bibitem{Bhattacharya:2016mcc}
B.~Bhattacharya, A.~Datta, J.~P.~Gu\'evin, D.~London and R.~Watanabe, Simultaneous explanation of the $R_K$ and $R_{D^{(*)}}$ puzzles: a model analysis, JHEP \textbf{01}, 015 (2017)
[arXiv:1609.09078 [hep-ph]].

\bibitem{Kumar:2018kmr}
J.~Kumar, D.~London and R.~Watanabe, Combined Explanations of the $b \to s \mu^+ \mu^-$ and $b \to c \tau^- {\bar\nu}$ Anomalies: a General Model Analysis,
Phys. Rev. D \textbf{99}, no.1, 015007 (2019)
[arXiv:1806.07403 [hep-ph]].

\bibitem{Gomez:2019xfw}
J.~D.~G\'omez, N.~Quintero and E.~Rojas, Charged current $b \to c \tau \bar{\nu}_\tau$ anomalies in a general $W^\prime$ boson scenario, Phys. Rev. D \textbf{100}, no.9, 093003 (2019)
\href{http://arxiv.org/abs/1907.08357}{[arXiv:1907.08357 [hep-ph]]}.

\bibitem{Iguro:2018fni}
S.~Iguro, Y.~Omura and M.~Takeuchi, Test of the $R(D^{(*)})$ anomaly at the LHC,
Phys. Rev. D \textbf{99}, no.7, 075013 (2019)
[arXiv:1810.05843 [hep-ph]].

\bibitem{Marzocca:2020ueu}
D.~Marzocca, U.~Min and M.~Son, Bottom-flavored mono-tau tails at the LHC, JHEP \textbf{12}, 035 (2020)
[arXiv:2008.07541 [hep-ph]].


\bibitem{Hayreter:2019dzc}
A.~Hayreter, X.~G.~He and G.~Valencia, LHC constraints on $W^\prime ,~Z^\prime $ that couple mainly to third generation fermions, Eur. Phys. J. C \textbf{80}, no.10, 912 (2020)
\href{http://arxiv.org/abs/1912.06344}{[arXiv:1912.06344 [hep-ph]]}.
  
\bibitem{He:2012zp} 
X.~G.~He and G.~Valencia, $B$ decays with $\tau$ leptons in nonuniversal left-right models, Phys.\ Rev.\ D {\bf 87}, 014014 (2013)   [arXiv:1211.0348 [hep-ph]].

\bibitem{He:2017bft}
X.-G. He and G. Valencia, Lepton universality violation and right-handed currents in $b \to c \tau \nu$,  Phys.\ Lett.\ B {\bf 779}, 52 (2018) \href{http://arxiv.org/abs/1711.09525}{arXiv:1711.09525  [hep-ph]}.

\bibitem{Carena:2018cow} 
  M.~Carena, E.~Meg\'{i}as, M.~Qu\'{i}ros and C.~Wagner, $ {R}_{D^{\left(*\right)}} $ in custodial warped space,  JHEP {\bf 1812}, 043 (2018) [arXiv:1809.01107 [hep-ph]].

\bibitem{Babu:2018vrl} 
K.~S.~Babu, R.~N.~Mohapatra and B.~Dutta, A Theory of $R(D^*,D)$ Anomaly with Right-Handed Currents,   JHEP {\bf 1901}, 168 (2019) [arXiv:1811.04496 [hep-ph]].
  
\bibitem{Asadi:2018wea}  
  P.~Asadi, M.~R.~Buckley and D.~Shih, It’s all right(-handed neutrinos): a new W$^{\prime}$ model for the $ {R}_{D^{{\left(\ast \right)}}} $ anomaly, JHEP {\bf 1809}, 010 (2018) [arXiv:1804.04135 [hep-ph]].  

\bibitem{Greljo:2018ogz} 
  A.~Greljo, D.~J.~Robinson, B.~Shakya and J.~Zupan, R(D$^{(\ast)}$) from $W^{\prime}$ and right-handed neutrinos, JHEP {\bf 1809}, 169 (2018)  [arXiv:1804.04642 [hep-ph]].

\bibitem{Robinson:2018gza} 
D.~J.~Robinson, B.~Shakya and J.~Zupan, Right-handed Neutrinos and $R(D^{(*)})$, JHEP {\bf 1902}, 119 (2019) [arXiv:1807.04753 [hep-ph]].

\bibitem{Asadi:2018sym} 
P.~Asadi, M.~R.~Buckley and D.~Shih, Asymmetry Observables and the Origin of $R_{D^{(*)}}$ Anomalies, Phys.\ Rev.\ D {\bf 99}, 035015 (2019) arXiv:1810.06597 [hep-ph].
  

\bibitem{Azatov:2018kzb} 
A.~Azatov, D.~Barducci, D.~Ghosh, D.~Marzocca and L.~Ubaldi, Combined explanations of B-physics anomalies: the sterile neutrino solution, JHEP {\bf 1810}, 092 (2018)  [arXiv:1807.10745 [hep-ph]].

\bibitem{Cvetic:2017gkt} 
G.~Cveti\v{c}, F.~Halzen, C.~S.~Kim and S.~Oh,  Anomalies in (semi)-leptonic $B$ Decays $B^{\pm} \to \tau^{\pm} \nu$, $B^{\pm} \to D \tau^{\pm} \nu$ and $B^{\pm} \to D^* \tau^{\pm} \nu$, and possible resolution with sterile neutrino, Chin.\ Phys.\ C {\bf 41}, 113102 (2017). \href{http://arxiv.org/abs/1702.04335}{[arXiv:1702.04335 [hep-ph]]}


\bibitem{Aloni:2017eny}
D.~Aloni, A.~Efrati, Y.~Grossman and Y.~Nir, $\Upsilon$ and $\psi$ leptonic decays as probes of solutions to the $R_D^{(*)}$ puzzle, JHEP \textbf{06}, 019 (2017)
[arXiv:1702.07356 [hep-ph]].

\bibitem{delAmoSanchez:2010bt}
P.~del Amo Sanchez \textit{et al.} [BaBar], Test of lepton universality in $\Upsilon(1S)$ decays at BaBar,
Phys. Rev. Lett. \textbf{104}, 191801 (2010)
[arXiv:1002.4358 [hep-ex]].

\bibitem{Besson:2006gj}
D.~Besson \textit{et al.} [CLEO], First Observation of $\Upsilon(3S) \to \tau^+ \tau^-$ and Tests of Lepton Universality in Upsilon Decays, Phys. Rev. Lett. \textbf{98}, 052002 (2007)
[arXiv:hep-ex/0607019 [hep-ex]].

\bibitem{Lees:2020kom}
J.~P.~Lees \textit{et al.} [BaBar], Precision measurement of the ${\cal B}(\Upsilon(3S)\to\tau^+\tau^-)/{\cal B}(\Upsilon(3S)\to\mu^+\mu^-)$ ratio, Phys. Rev. Lett. \textbf{125}, 241801 (2020)
[arXiv:2005.01230 [hep-ex]].

\bibitem{Amhis:2016xyh}
Y.~Amhis \textit{et al.} [HFLAV], Averages of $b$-hadron, $c$-hadron, and $\tau$-lepton properties as of summer 2016, Eur. Phys. J. C \textbf{77}, no.12, 895 (2017)
[arXiv:1612.07233 [hep-ex]].


\bibitem{Kamali:2018bdp} 
S.~Kamali, New physics in inclusive semileptonic $B$ decays including nonperturbative corrections, Int.\ J.\ Mod.\ Phys.\ A {\bf 34}, 1950036 (2019) [arXiv:1811.07393 [hep-ph]].

\bibitem{Kou:2018nap} 
E.~Kou {\it et al.} [Belle-II Collaboration], The Belle II Physics Book, PTEP {\bf 2019}, no. 12, 123C01 (2019)
Erratum: [PTEP {\bf 2020}, no. 2, 029201 (2020)]
[arXiv:1808.10567 [hep-ex]].
  
\bibitem{Cardozo:2020uol}
J.~Cardozo, J.~H.~Mu\~noz, N.~Quintero and E.~Rojas, Analysing the charged scalar boson contribution to the charged-current $B$ meson anomalies, J. Phys. G \textbf{48}, no.3, 035001 (2021)
[arXiv:2006.07751 [hep-ph]].

\bibitem{Garcia:2021}
C. H. Garc\'{i}a-Duque, J.~H.~Mu\~noz, N.~Quintero and E.~Rojas, Work in progress (2021).
\end{thebibliography}
\end{document}